\documentclass[english,prl,A4,twocolumn,english,showpacs,aps]{revtex4}
\usepackage{graphicx}
\usepackage{amsmath}
\usepackage{amssymb}
\usepackage{epstopdf}

\begin{document}
\newcommand{\bi}{\boldsymbol}
\newcommand{\Fref}{\ref}
\newcommand{\ApJ}{ApJ}
\newcommand{\ApJL}{ApJ Lett.}
\newcommand{\ApJS}{ApJ Suppl.}
\newcommand{\PRL}{Phys. Rev. Lett.}
\newcommand{\PRD}{Phys. Rev. D}
\newcommand{\PRE}{Phys. Rev. D}
\newcommand{\MNRAS}{MNRAS}
\newcommand{\mnras}{MNRAS}
\newcommand{\jcph}{J.Comp.Phys.}
\newcommand{\ARAA}{ARA\&A}
\newcommand{\AsAs}{A\&A}
\newcommand{\PASP}{PASP}
\newcommand{\pasp}{PASP}


\title{\textbf{The influence of gas-kinetic evolution on plasma reactions}}
\author{D A Diver\email{bla}, L F A Teodoro, C S MacLachlan and H E Potts}
\affiliation{Dept of Physics and Astronomy, University of Glasgow, Glasgow G12 8QQ, UK}
\email[email: ]{d.diver@physics.gla.ac.uk}
\date{\today}

\begin{abstract}
Plasmas in which there is a threshold for a dominant reaction to take place (such as recombination or attachment) will have particle distributions that evolve as the reaction progresses. The form of the Boltzmann collision term in such a context will cause the distribution to drift from its initial form, and so cause for example temperature fluctuations in the plasma if the distribution is originally Maxwellian. This behaviour will impact on the relevant reaction rates in a feedback loop that is missing from simple chemical kinetic descriptions since the plasma cannot be considered to be isothermal, as is the case in the latter approach. In this article we present a simple kinetic model that captures these essential features, showing how
cumulative differences in the instantaneous species levels can arise over the purely chemical kinetic description, with implications for process yields and efficiencies.
\end{abstract}
\pacs{52.20.Fs, 52.20.Hv, 52.25.Dg, 52.25.Gj, 52.25.Ya, 52.27.Gr, 82.33.Xj}
\maketitle

\section{Introduction}
In many applications and contexts plasmas that evolve to a critical non-equilibrium stage are vitally important, particularly where such plasmas develop to a stage where either an instability is triggered (such as the onset of a set of reactions that alter the plasma characteristics) or a significant phase change is induced (the plasma evolves to a substantially different equilibrium condition). Examples of such behaviour are when plasmas become significantly electronegative \citep[for example][]{corr03, lieb1, lieb2} or quench completely to form neutral gas \citep[for example][]{tanaka, sentman, janev, puy, benetti}.
The purpose of this article is to explore the implications in the gas-kinetic framework of plasmas in which the distribution function of the electrons is altered by unbalanced plasma chemistry; reactions with strong electron energy dependence such as electron recombination with positive ions to form neutrals, or electron attachment to neutrals to form negative ions, can lead to significant changes in the electron energy distribution function (eedf) which in turn affect the progress of subsequent reactions.

Very often the modelling of plasma reactions is implemented in the fluid context, using rate coefficients to determine the evolution of fluid plasma species. However there is an implicit assumption in such fluid approaches, namely that the underlying plasma distribution is Maxwellian, with a well-defined temperature; no macroscopic variables other than the density (the zeroth kinetic moment) enter such equations, and the temperature therefore is an externally defined parameter of the problem. The rate coefficients are often presented in Arrhenius-type formulae that make the temperature dependence explicit, and the reaction set is then modelled by rate equations which are primarily driven by species number density variations within the Maxwellian context. For example \cite{Lieb:2005, IUPAC:1997, sentman, mendez},
\begin{equation}
k=A(T/T_0)^n\exp[-E_a/(k_BT)] \label{arrh}
\end{equation}
is the general Arrhenius form for a chemical process with activation energy $E_a$, and reference temperature $T_0$; $n$ is an additional parameter, theoretically predicted to take the value $1/2$.  
 An explicit example is that for the dissociative electron attachment to an ozone molecule, forming a negative atomic ion and a neutral molecule \cite{1999APS..GEC.AT106S, 0022-3727-37-15-005}
\begin{eqnarray}
\left \{
\begin{array}{c}
\mbox{e + O}_3\rightarrow \mbox{O}_2+\mbox{O}^-  \\
\\
k=2.12 \times 10^{-15}T_e^{-1.06}\exp(-0.93/T_e)\,\mbox{m}^{3}\mbox{s}^{-1}\label{ozoneatt}
\end{array} \right .
\end{eqnarray}
in which $T_e$ is given in electronvolts. In this article we explore the possibility that the plasma reactions affect the plasma temperature (more correctly, the mean energy per particle) in a feedback loop, simply because the distribution function of each species must be altered as a direct result of reactions converting one set of plasma species to another set. In order to describe this as a distribution function evolution, we require a gas-kinetic description based on the full reaction cross-sections, not to be confused with a chemical-kinetic one which uses the fluid-based rate constant, and which is an averaged treatment. In general, the rate constant $k$ used in chemical kinetic descriptions is derived from the gas-kinetic reaction cross-section $\sigma$ by the following integration over velocity space $\bi{v}$ \cite{Lieb:2005}:
\begin{equation}
k=\frac{\int \sigma(\bi{v}) f(\bi{v},t) v\, \mbox{d} \bi{v}}{\int f(\bi{v},t) \,\mbox{d} \bi{v}}\label{kformal}
\end{equation}
where $f(\bi{v},t)$ is the distribution function of the species (taken as Maxwellian) implicated in the reaction associated with the cross-section $\sigma$, $\bi{v}$ is the velocity co-ordinate, and the integration is over the whole velocity space. The casting of the formal expression for $k$ from Eq.~(\ref{kformal}) in the (modified) Arrhenius form of Eq.~(\ref{arrh}) is generally done by an empirical fit to experimental data, valid over a limited energy range.

In this article we will concentrate on electron-moderated reactions, but the principle applies to all species. For instance, electron attachment and detachment can have a significant effect on the electron distribution within a plasma; another example is recombination, in which electrons and positive ions recombined to form neutrals, simultaneously depleting the positive and negative species whilst augmenting the neutrals. The fact that such reactions proceed according to a reaction cross-section in kinetic theory means that the evolution of the distribution function of each species can be accommodated in such a model; however, modelling in the fluid context with rate equations averages out distribution function effects by requiring that the underlying distribution is always Maxwellian for rate coefficients that are temperature dependent. The sections following this one describe the background kinetic theory, and show how the simplest possible model encompasses a temperature drift arising directly from the influence of the reactions on the form of the electron distribution function.
\section{Background Theory}
Consider a plasma in which binary interactions maintain the equilibrium.
Then the distribution function $f_s$ for a given species $s$ obeys Boltzmann's equation,
\begin{equation}
\frac{\partial  f_s}{\partial t}+\bi{u}\cdot \frac{\partial f_s}{\partial \bi{r}}+\bi{a_s}\cdot \frac{\partial f_s}{\partial \bi{u}}=\left(\frac{\partial f_s}{\partial t}\right)_{\!\!c}
\end{equation}
where $\bi{a}_s$ is the acceleration, $\bi{r}$ is the spatial co-ordinate, and $\bi{u}$ is the velocity co-ordinate. The distribution function is defined such that $f(\bi{r}, \bi{u},t)\mbox{d}\bi{u}$ gives the number of particles in the velocity range $\bi{u}$ to $\bi{u}+\mbox{d}\bi{u}$ at the position $\bi{r}$ at time $t$.
The collision term on the right gives the change in the distribution arising from binary collisions, and is given by Boltzmann's formula \cite{dadbook}:
\begin{eqnarray}
&&\left(\frac{\partial f_s}{\partial t}\right)_{\!\!c}=\nonumber\\
 &&\quad \,\sum_{j}\int\left[ f_s(\bi{r},\bi{u}',t)f_j(\bi{r},\bi{u}'_j,t)-f_s(\bi{r},\bi{u},t)f_j(\bi{r},\bi{u}_j,t)\right] \nonumber \\
 & &\quad \times~|\bi{u}-\bi{u}_j|\,\sigma_{js} \mbox{d}\bi{u}_j \label{bci}
\end{eqnarray}
where $f_j$ denotes the distribution function of the target particles, and $'$ denotes post-collision quantities; $\sigma_{js}$ is the collision cross-section for the interaction between particles $j$ and $s$. The term in square brackets shows how the distribution function is changed by the flux of particles scattered into the velocity range $\bi{u}$ to $\bi{u}+\mbox{d}\bi{u}$ minus those scattered out of that range, as a result of the interaction governed by the cross-section $\sigma_{sj}$. Note that a kinetic plasma in equilibrium has zero collision term. The kinetic theory of plasmas will yield bulk (that is, macroscopic or fluid) quantities by integrating over all velocity space. For example, we can define the number density $n(\bi{r},t)$ and mean square speed per particle $\langle u^2 \rangle (\bi{r},t)$ as moments of the distribution function:
\begin{eqnarray}
n(\bi{r},t)&=&\int f(\bi{r},\bi{u},t)\,\mbox{d}\bi{u} \\
\langle u^2\rangle (\bi{r},t)&=&\int u^2f(\bi{r},\bi{u},t)\,\mbox{d}\bi{u}\left[\int f(\bi{r},\bi{u},t)\,\mbox{d}\bi{u}\right]^{-1}
\end{eqnarray}
In a Maxwellian equilibrium, the mean energy per particle defines the temperature $T$:
\begin{equation}
\case{1}{2}m\langle u^2 \rangle =\case{1}{2}N_dk_BT
\end{equation}
where $m$ is the particle mass, and $N_d$ is the number of spatial degrees of freedom. Very often the electron distribution is not actually Maxwellian, but the concept of an effective temperature is deduced from the mean energy calculated from the second moment in this way.

The Boltzmann collision integral Eq.~(\ref{bci}) has restricted validity: it is only a good description of binary, uncorrelated interactions which take place over length and time scales much shorter than any other intrinsic variation in $f$ itself (such as plasma inhomogeneity or external forcing). Since charged particles in a fully ionized plasma tend to interact collectively via simultaneous long-range Coulomb forces arising from the disposition of other charges, the binary interaction is less appropriate, and the Boltzmann collision term is often replaced with the more sophisticated Fokker-Planck treatment. However for a weakly ionized plasma in which such many-body Coulomb interactions do not dominate over binary encounters with neutral species or other charged particles, the Boltzmann description is still very relevant.
\section{Simple model: weakly ionized plasma}
Consider the simple case of a weakly ionized kinetic plasma, governed by the Boltzmann collision integral, in which a particular reaction suddenly starts in an unbalanced way. For example, this could be an electronegative plasma in which electron attachment becomes the dominant driving term under a set of prevailing local conditions, or perhaps a plasma begins to quench as a result of recombination becoming important as a local energy density drops. In each scenario, the key issue is that the reaction is initiated in a plasma in such a way that it is predominantly one-sided, that is, it is far from equilibrium for that reaction. For the sake of being specific, consider a plasma in which the temperature has fallen sufficiently that recombination suddenly becomes a significant reaction. The sequence of events we imagine to proceed as follows:
\begin{enumerate}
\item The plasma cools below some critical energy threshold such that the balance between recombination and ionization is significantly altered to favour the former
\item Binary recombination dominates, leading to a loss of electrons and positive ions according to the energy dependence of the cross-section
\item The distribution function evolves according to the Boltzmann equation, with the collision integral explicitly yielding corrections to the distribution function on timescales shorter thn the equilibration time
\item The distribution function equilibrates to a new mean energy (or temperature, if Maxwellian) and the reactions proceed according to the new, evolved form of the distribution function that governs the remaining charged particles
\end{enumerate}

The significance of this description lies in the fact that the mean energy per particle (proportional to the temperature, if Maxwellian) is evolving at the same time as the reaction proceeds, since if we consider the corrections $\delta f$ to the distribution function at time $t$ arising from the loss of particles, then the instantaneous distribution function will not take the same form as the initial one: the form of the collision integral Eq.~(\ref{bci}) precludes this for all but pathological forms for $\sigma$. Relaxation back to a Maxwellian (for example) for fewer particles may well be possible, but will necessarily change the temperature, since there has to be a redistribution of particles in velocity (and energy) space.

In kinetic terms, this behaviour can be readily accommodated (in principle); however, in the fluid context this is problematical, since there is an implicit assumption that the distribution function is always in its equilibrium form, that is, Maxwellian. Fluid models therefore use rate constants, rather than reaction cross-sections, and so the feedback loop that connects the progress of the reactions to the evolving form of the participating distribution functions is broken. Non-equilibrium effects in plasma chemistry are known to be significant \cite{kruger89, kruger97} and the reconstruction of effective reaction rates from gas-kinetic simulations can provide information inaccessible from only fluid models \cite{tong}.

If as a direct result of kinetic effects either the temperature drifts, or if the distribution functions of the reactants become non-Maxwellian, then the assumed isothermal rate coefficients may be inaccurate, leading to cumulative discrepancies in the relative numbers of species.

The simplest case we can do for illustrative purposes is the 
1-velocity dimension kinetic plasma in which a particular reaction becomes predominantly one-way, leading to electron losses (that is, the electrons are recombining with positive ions or perhaps are attaching to neutrals). Let's assume that the electrons and ions (or neutrals) (denoted by subscripts $e$, $i$ and $n$ respectively) are Maxwellian initially, with temperature $T_0$, though the analytical framework is more general than this. Since the reaction is essentially one-sided, meaning that electrons are much more likely to be captured (and lost to the distribution) than scattered into a different velocity range, we can approximate the collision term by retaining only the part that describes the loss of electrons from a velocity element:
\begin{equation}
\left(\frac{\partial f_e}{\partial t}\right)_{\!\!c} 
\approx-\int_{-\infty}^{\infty}f_ef_s|v_i-v_e|\sigma\,\mbox{d}v_i\label{kinloss}
\end{equation}
where $s$ denotes the species with which the electron is interacting ($s=i$ for recombination, $s=n$ for electron attachment to neutral species $n$, etc), and we have simplified the notation by dropping functional arguments. The  cross-section for the dominant process under consideration is denoted simply by $\sigma$. Of course, should the reaction be creating more free electrons, then the sign of the right-hand side of Eq.~(\ref{kinloss}) is reversed. Note that there is also a self-term for the electrons that will lead to a rearrangement of the electron distribution without changing the total number of electrons; the importance of this self-relaxation term has to be judged in terms of the relative timescale for the competing reaction. We will first evaluate the effect of the electron loss on the electron distribution function as a result of the continuing reaction, and then address how this may be modified by relaxation processes, given that electron-electron relaxation is best treated as a simultaneous, collective interaction rather than a sequence of binary interactions.

In order to better illustrate the issue at the heart of this article, we will move from what has been generally valid to a specific case of Maxwellian electrons. Of course, other driven equilibria are perfectly possible, and can be described by the same theoretical framework, but the integrals are known and tractable for the Maxwellian plasma, and so we will report the results in this case.

Taking Maxwellians at a common temperature $T$ initially for the electrons and ions,
\begin{eqnarray}
f_s&=&C_s e^{-\alpha_sv_s^2}\\
\alpha_s&=&m_s/(2k_BT_0)\\
C_s &=& n_{0s}\sqrt{\alpha_s/\pi}
\end{eqnarray}
in which $s=i$ or $e$, and $n_{0s}$ is the initial species number density.
We are approximating the cross-section as a step function in energy, such that $\sigma=\sigma_0$ for all speeds less than some critical speed (to be defined later), and $\sigma=0$ otherwise. (In practice, if the critical speed is greater than the thermal speed, there is a negligible contribution to the integral of a non-zero cross-section at values higher than the critical speed for a Maxwellian distribution; hence taking the cross-section as constant for all energies doesn't significantly change the results.)  The simplicity of this appproximation is that all of the key integrals can be done analytically, illustrating the concept. The Appendix gives the full form of the integrals; to minimise the disruption to the flow of the argument we will simply state the analytical results here:

\noindent\begin{eqnarray}
\int_{-\infty}^{\infty}f_ef_i|v_i-v_e|\sigma_0\,\mbox{d}v_i \nonumber ~~~~~~~~~~~~~~~~~~~~~~~~~~~~~~~~~~~
\end{eqnarray}
\vspace{-0.5cm}
\begin{eqnarray}
&=& \sigma_0C_eC_ie^{-\alpha_ev_e^2} 
\int_{-\infty}^{\infty}e^{-\alpha_i v_i^2}|v_i-v_e|\,\mbox{d}v_i\\
&=&n_{0i}\sigma_0C_e|v_e| e^{-\alpha_ev_e^2}\mbox{erf}(\sqrt{\alpha_i}|v_e|)\nonumber\\
&&+\sigma_0n_{0i}n_{0e}\pi^{-1}\sqrt{\alpha_e/\alpha_i}\exp[-(\alpha_i+\alpha_e)v_e^2]\label{firstint}\\
&=&-\left(\frac{\partial f_e}{\partial t}\right)_{\!\!c}
\end{eqnarray}
This yields the total rate of change of $f_e$ resulting from the recombination process, leading to the evolved distribution \begin{eqnarray}
f_e(\Delta t)&\approx& f_e(0)+\Delta t\left(\frac{\partial f_e(0)}{\partial t}\right)_{\!\!c}\nonumber\\
&=&C_ee^{-\alpha_ev_e^2}\left\{1-n_0\sigma_0\Delta t \left[|v_e|\,\mbox{erf}(\sqrt{\alpha_i}|v_e|) \right . \right . \nonumber \\
& &  + \left. \left. (\pi\alpha_i)^{-1/2}e^{-\alpha_iv_e^2}\right]\right\}\label{fdeltat}
\end{eqnarray}
In order to maintain a physically meaningful corrected distribution, we need $f_e(\Delta t)\ge 0$, and so we must restrict the validity of (\ref{fdeltat}) to values of $v_e$ such that
\begin{equation}
n_0\sigma_0\Delta t|v_e| \lesssim 1 \label{velrestrict}
\end{equation}
bearing in mind that $\alpha_i \gg \alpha_e$, and so the value of the error function can be taken to be unity, and neglecting the last term on the right-hand side of Eq.~(\ref{fdeltat}), which is not significant at finite $|v_e|$. This restriction can be accommodated assuming that an energy (or indeed velocity) dependent cross-section will fall sharply with energy (or speed) to become negligible in value outside a critical energy range, so that it closely resembles a step-function (for example, some electron attachment reactions are only significant below a critical threshold energy); the formal statement of Eq.~(\ref{velrestrict}) stems from the fact that we hold the cross-section to be constant for all speeds. If instead we rephrase this to be $\sigma=\sigma_0$ for $|v_e|<1/(n_0\sigma_0\Delta t)$; $\sigma=0$ otherwise, then we have a preliminary basis for defining the critical speed mentioned earlier, and therefore formalising the approach.

\begin{figure}[!]
{\scalebox{0.68}{\includegraphics{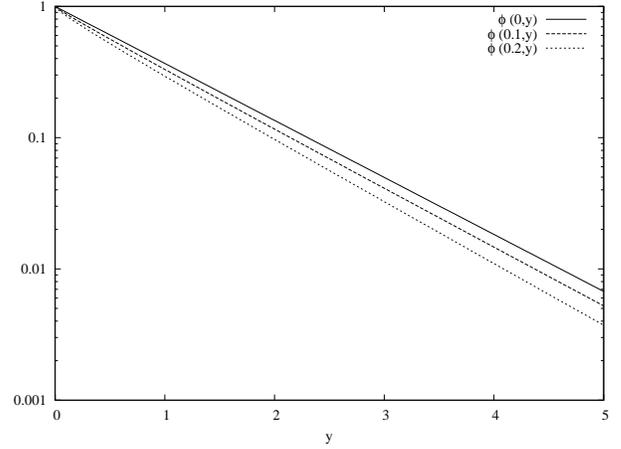}}}
\caption{\label{pplot} Plot of the function $\phi(a,y)$ from (\ref{psifull}) as a function of normalised energy $y$ for $a=0$ (pure Maxwellian), $a=0.1$ and $a=0.2$, illustrating the drift from the pure Maxwellian as a function of collision correction. Note that non-zero values of $a$ still give the appearance of a straight-line solution in the log-linear plot, apart from significant curvature at low energies, though clearly the inferred temperature is lower than the true Maxwellian case. }
\end{figure}

We can now ask the question: is the evolved distribution given in Eq.~(\ref{fdeltat}) a Maxwellian?  The answer is no, since if we take the function
\begin{equation}
\phi(a,y)=e^{-y}(1-ay^{1/2})\label{psifull}
\end{equation}
as a simplified, scaled approximation of $f_e(\Delta t)$ expressed in terms of energy, rather than speed, we can see that
\begin{equation}
\log \phi(a,y)\approx -y-a\sqrt{y} \label{psi}
\end{equation}
A true Maxwellian would have only the first term on the right-hand side in Eq.~(\ref{psi}); there is an additional small, non-zero curvature correction to the straight line expected from the log-linear plot, characterising the drift from a Maxwellian as a result of the unbalanced binary interaction. Figure \ref{pplot}  gives a simple illustration of this effect. Log-linear plots of the distribution function as a function of energy with different values of the collision correction show that the latter increases the apparent gradient of the plot, showing that the inferred `temperature' (proportional to the reciprocal gradient) is lower than that associated with the pure Maxwellian. This change from a Maxwellian shouldn't be too surprising, given the nature of the binary collision term in Eq.~(\ref{kinloss}): the distributions of the electrons and target particles are involved in a (relative) velocity-scaled convolution product, and only pathological cases could yield a gaussian form.

In order to quantify the phenomenological consequences of this correction, we can calculate key macroscopic observables.
By integrating Eq.~(\ref{fdeltat}) over electron velocity space we can evaluate the accompanying rate of change of electron number density:
\begin{equation}
\frac{\partial n_e}{\partial t}=-n_{0e}n_{0i}\sigma_0\pi^{-1/2}\left(\frac{\alpha_i+\alpha_e}{\alpha_i\alpha_e}\right)^{1/2}
\end{equation}
To get the change in the mean energy of the particles, we need to multiply Eq.~(\ref{firstint}) by $v_e^2$ and integrate again over electron velocity space, resulting in the expression
\begin{equation}
\frac{\partial \langle n_ev_e^2\rangle }{\partial t}=
-\case{1}{2}\pi^{-1/2}n_{0e}n_{0i}\sigma_0\left(\frac{\alpha_i\alpha_e}{\alpha_i+\alpha_e}\right)^{1/2}\left(\frac{2\alpha_i+\alpha_e}{\alpha_i\alpha_e^2}\right)
\end{equation}

Hence we can approximate the density and energy evolution in time $\Delta t$ as follows:
\begin{eqnarray}
n_e(\Delta t)&\approx& n_{0e}+\left(\frac{\partial n_e}{\partial t}\right)_{\!\!c}\Delta t\\
&=& n_{0e}-\pi^{-1/2}n_{0e}n_{0i}\sigma_0\left(\frac{\alpha_i+\alpha_e}{\alpha_i\alpha_e}\right)^{1/2}\label {ndot}\\[9pt]
\langle n_ev_e^2 \rangle (\Delta t)&\approx& \langle n_{0e}v^2(0)\rangle +\left(\frac{\partial \langle n_ev_e^2\rangle}{\partial t}\right)_{\!\!c}\Delta t\\[12pt]&=&
\frac{n_{0e}}{2\alpha_e}-\frac{n_{0i}n_{0e}\sigma_0}{2\sqrt{\pi}}\left(\frac{\alpha_i\alpha_e}{\alpha_i+\alpha_e}\right)^{1/2} \nonumber \\
&  & \times \left(\frac{2\alpha_i+\alpha_e}{\alpha_i\alpha_e^2}\right) \Delta t \label{nvsqdot}
\end{eqnarray}
so long as the time interval $\Delta t$ is sufficiently short that the distribution function has not changed significantly, that is, the perturbation in particle number density must be small.

In order to interpret the change in the mean energy as a change in temperature, we need to consider the equivalent Maxwellian distribution that has the same particle number density and total energy as the collision-amended one. The temperature $T_1$ of this equivalent Maxwellian can be given in terms of the original one $T_0$ by considering the ratio of the second moment to the zeroth of the post-collision distribution function, Taylor expanded in time using equations\,(\ref{ndot}) and (\ref{nvsqdot}). This yields
\begin{equation}
\frac{T_1}{T_0}=1-\frac{\Delta T}{T_0}= 1-n_{0i}\sigma_0\Delta t\left[\frac{\alpha_i}{\pi\alpha_e(\alpha_i+\alpha_e)}\right]^{1/2} \label{tdrift}
\end{equation}
showing that the temperature drops as the interaction causes particles to be lost to the distribution, consistent with the simple picture presented in Figure \ref{pplot}.

Now we can quantify the permitted time interval for this approximation to be valid: the second term in Eq.~(\ref{tdrift}) must be small, yielding an appropriate value for $\Delta t$, the distribution function evolution time. If this $\Delta t$ is similar to the equilibration time, then it is reasonable to assume that the plasma will relax to a new Maxwellian in $\Delta t$, giving a formal statistical basis for the evolving temperature. However, the relaxation might not be rapid in a weakly ionized plasma, since the electrons and ions are dominated by neutrals, and therefore the relatively rapid Coulomb interaction which would equilibrate a fully ionized plasma may in fact be dominated by electron-neutral collisions in such a strongly-coupled plasma (a plasma is weakly coupled if the kinetic energy of the particles greatly exceeds any local potential energy, and strongly coupled if the kinetic energy is dominated by local potentials).

Hence there are several possibilities for the evolution of the plasma in time $\Delta t$:
\begin{itemize}
  \item the electrons and the positive ions could both relax to a new Maxwellian with a new temperature;
  \item the electrons might relax to a Maxwellian, but the ions remain non-Maxwellian;
  \item neither species relaxes to a Maxwellian
\end{itemize}
Note that the relaxation of electrons must be via interaction with species other than the target one implicated in the reaction, since interaction with the latter is demonstrated to produce an evolution away from a Maxwellian. The most rapid relaxation process is self-relaxation, given the superior mobility of electrons compared to any other species present, and this process is the most likely to produce a Maxwellian form for the electrons.

In the first two cases, Eq.~(\ref{tdrift}) is an approximate Taylor-series expansion of the electron temperature evolution with time. In order to quantify the magnitude of this temperature change, it is necessary to specify a value for $\Delta t$. Since the assumption underlying Eq.~(\ref{tdrift}) is that the electrons have relaxed to a Maxwellian, it is appropriate to take the time interval to be equal to the self-equilibration time $\tau_E^{\mbox{\tiny{self}}}$ for electrons \cite{dadbook, Boyd:1969}:
\begin{eqnarray}
\Delta t &=&\tau_E^{\mbox{\tiny{self}}}\nonumber\\
&\approx& \frac{(2k_BT/m_e)^{3/2}}{4\alpha_r\Psi(1)}\label{tau_e}\\
\alpha_r&=& \frac{e^4n_{0e}\ln \Lambda}{2\pi\epsilon_0^2m_e^2}\\
\Psi(x)&=&\frac{\mbox{erf}(x)-x\,\mbox{erf}\,'(x)}{2x^2}
\end{eqnarray}
where $\ln\Lambda$ is the usual Coloumb logarithm.

Recognising that $\alpha_e \ll \alpha_i$, we can simplify the expression for the fractional temperature change:
\begin{eqnarray}
\frac{\Delta T}{T_0} &\approx& \frac{\sigma_0}{4\Psi(1)\ln \Lambda}\frac{2\sqrt{\pi}\epsilon_0^2(2k_BT_0)^2}{e^4}\nonumber\\
&\approx& 3\times 10^8\times \frac{\sigma_0 T_0^2}{\ln \Lambda}\label{tfracchange}
\end{eqnarray}
where we have taken $4\Psi(1)\approx 1$. The Coulomb logarithm is usually taken to lie in the range $2 \le \ln\Lambda \le 20$, with the upper limit associated with fully ionised, weakly coupled plasmas, and the lower one with partially ionised, strongly-coupled plasmas, though some caution has to be exercised here: the validity of such assumptions has been challenged for plasmas that are not the idealised, Maxwellian ones for which the Fokker-Planck analysis holds perfectly (and therefore the concept of self-relaxation producing a $\ln \Lambda$ term is consistent) \cite{zha:2006, 2001NucFu..41..631L},
and in reality, the value of $\ln \Lambda$ may be only half of the classical one.

For a relevant specific example, let us return to the example of electron attachment to ozone, for which the rate constant in Arrhenius form was given in Eq.~(\ref{ozoneatt}). The cross-section for this dissociative process has a peak at around 1 eV \cite{1999APS..GEC.AT106S,1997PSST....6..140M}
(there is a resonance below 1 eV which we shall disregard in the context of a plasma cooling to the process threshold at around 3 eV). The peak value of the cross-section (excluding the resonance) is $\sim 3\times 10^{-21}$m$^2$. If we take $\sigma_0 \sim 2\times 10^{-21}$m$^2$, and take $\ln\Lambda \sim 2$, $T_0\sim 3\times 10^4$K, then we find from Eq.~(\ref{tfracchange}) that
\begin{equation}
\frac{\Delta T}{T_0}\sim 3\times 10^{-4}
\end{equation}
so that the fractional temperature change is under $0.1\%$ on an
electron equilibration time. A plasma with a significant ozone
fraction is clearly only partially ionized, and so the validity of
using the equilibration time Eq.~(\ref{tau_e}) is open to question.
Nevertheless, this temperature drift is cumulative, since cooling
the electrons maintains the likelihood of the reaction. Thus in the
fluid context, the minimum temperature fluctuation must be around
1-2 orders of magnitude greater on the smallest fluid timescale,
since the validity of the fluid approximation requires many
collision times to elapse during the smallest resolvable fluid time.
The significance of this lies in the evolution of the rate constant,
which is a fluid quantity. The fractional rate of change of $k$ in
Arrhenius form Eq.~(\ref{arrh}) is
\begin{equation}
\frac{\dot{k}}{k}=(n+E_a/k_BT)\frac{\dot{T}}{T}\label{kevol}
\end{equation}
showing that the fractional drift in the rate constant $k$ scales
with the fractional temperature drift as a function of $n$ and
$E_a$, as well as $T$. Hence for a system of fluid species evolution
equations, the drift in each rate constant can be different, leading
to cumulative changes in the evolving species populations compared to those
arising under the usual assumption of uniform temperature.

It is worth noting here that depending on the sign of the factor on the right-hand-side of Eq.~(\ref{kevol}), the rate constant changes may be in or out of phase with the temperature drift. Hence it may be possible for the temperature to drift below a minimum threshold for a reaction, causing it to stop temporarily pending additional local heating. Such instabilities would only be apparent in the kinetic limit, but could have consequences for the overall efficacy of a plasma reaction process.
\section{Illustrating the effect of kinetic feedback on the fluid model}
In order to explore the consequences of Eq.~(\ref{kevol}), consider the simple chemical kinetics reaction
\begin{equation}
\mbox{e + A}\rightarrow \mbox{B}\label{1sidedreaction}
\end{equation}
in which species A is modified by electron attachment or recombination to produce species B. Since the context here is the non-equilibrium evolution of the plasma, the reverse reaction has a negligible role to play. The rate equations for this reaction are
\begin{eqnarray}
\dot{n}_e &=& -k n_e n_{\mbox{\tiny{A}}}\nonumber\\
\dot{n}_{\mbox{\tiny{A}}} &=&-k n_e n_{\mbox{\tiny{A}}}\nonumber\\
\dot{n}_{\mbox{\tiny{B}}} &=&+k n_e n_{\mbox{\tiny{A}}} \label{reactionset}
\end{eqnarray}
where we are only taking equations for the one-sided reaction of Eq.~(\ref{1sidedreaction}), with $n_e$ the electron number density, $k$ the rate constant for the reaction and $n_{\mbox{\tiny{A}}}$, $n_{\mbox{\tiny{B}}}$ the number densities of species A and B respectively.

By eliminating $n_{\mbox{\tiny{A}}}$ from the first two equations in Eq.~(\ref{reactionset}), we arrive at a single equation for the evolution of the electron number density:
\begin{equation}
\frac{\mbox{d}}{\mbox{d}t}\left(\frac{\dot{n}_e}{k n_e}\right)=-\dot{n}_{\mbox{\tiny{A}}}=-\dot{n}_e
\end{equation}
Integrating once, and neglecting the integration constant, we have
\begin{equation}
\dot{n}_e+k n_e^2=0\label{eq:nointegconst}
\end{equation}
which has solution
\begin{equation}
n_e(t)=n_{e0}\left[ 1+n_{e0}\int_0^tk(t')\,\mbox{d}t'\right]^{-1}\label{ne_soln}
\end{equation}
where $n_e(0)=n_{e0}$ and we have allowed the possibility of $k$ being time-dependent, motivated by the explicit temperature dependence of rate equations shown in Eq.~(\ref{arrh}), and the fact that this temperature must drift in time, according to Eq.~(\ref{tfracchange}).

Since the equation for $n_{\mbox{\tiny{A}}}$ is identical to that for $n_e$, then species A evolves in time according to
\begin{equation}
n_{\mbox{\tiny{A}}}(t)=n_{\mbox{\tiny{A}}0}\left[ 1+n_{\mbox{\tiny{A}}0}\int_0^tk(t')\,\mbox{d}t'\right]^{-1}\label{nA_soln}
\end{equation}
with the change in species B being determined via
\begin{equation}
n_{\mbox{\tiny{A}}}+n_{\mbox{\tiny{B}}}=\mbox{constant}=n_0,\mbox{ say}\label{totaldens}
\end{equation}
which is evident from the sum of the last two equations in Eq.~(\ref{reactionset}). Note that the evolution of $n_e$ in Eq.~(\ref{ne_soln}) is constrained by Eq.~(\ref{totaldens}).

Modelling the time dependence of $k$ as
\begin{equation}
k=k_0+k_1t^p
\end{equation}
where $k_0$, $k_1$ and $p>0$ are constants, we can contrast the electron number density at time $t$ for the case $k_1=0$ (rate constant has fixed value, implying isothermal evolution) with $k_1\neq 0$ (time-dependent rate constant for a non-isothermal evolution) by defining the ratio $R(t)$ as follows:
\begin{eqnarray}
R(t)=\frac{n_e(t;k_1\neq 0)}{n_e(t;k_1=0)}&=&\frac{1+n_{e0}k_0t}{1+n_{e0}k_0t \Phi}\label{eq:neratio}\\[8pt]
&\approx&1+\beta(1-\Phi);\;\;\beta\ll1\\
&\approx& 1/\Phi ;\;\;\beta\gg1
\end{eqnarray}
where
\begin{equation}
\beta =n_{e0}k_0t
\end{equation}
is a characteristic time, and
\begin{equation}
\Phi=1+\frac{k_1}{k_0}\left(\frac{\beta}{n_{e0}k_0}\right)^p
\end{equation}

\begin{figure}[!]
{\scalebox{0.68}{\includegraphics{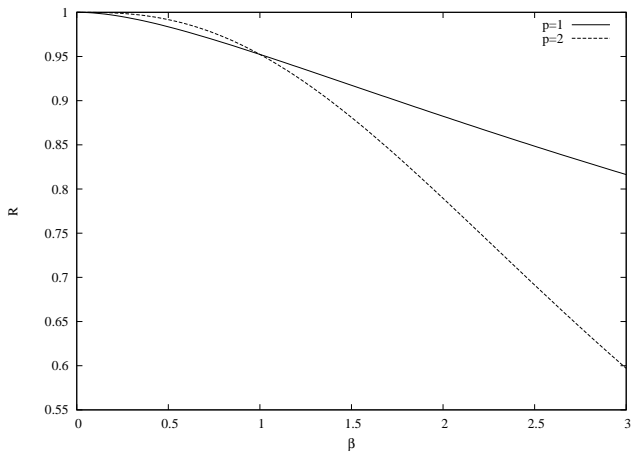}}}
\caption{\label{neplot} Plot of the function $R(\beta)$ defined in Eq.~(\ref{eq:neratio}) for $p=1$ and $p=2$, illustrating the increasing discrepancy in the electron number density with time for time-dependent rate coefficients compared to non-evolving rate constants. After 2 characteristic times, the discrepancy is around 10\% for $p=1$, and around $20\%$ for $p=2$.}
\end{figure}

Hence depending on the elapsed time, the population of species can be significantly different if the plasma is not isothermal as the reaction progresses. An illustration of this effect is shown in \Fref{neplot}, where the ratio $R(\beta)$ from Eq.~(\ref{eq:neratio}) is plotted for two cases: $p=1$ and $p=2$. In each case, the horizontal axis is the characteristic time $\beta$, and $\phi=1+0.1x^p$, simply for illustrative purposes. It is evident from the graphs that there can be a discrepancy in the electron number density of up to $20\%$ after a couple of characteristic times.

In Eq.~(\ref{eq:nointegconst}) we neglected an integration constant; restoring it yields a more general differential equation:
\begin{equation}
\dot{n}_e+k n_e^2=kCn_e
\end{equation}
with solutions in the form
\begin{equation}
n_e(t)=\frac{C}{1-(1-C/n_{e0})\exp{\left[-C\int_0^tk(t')\mbox{d}t'\right]}}\label{eq:genform}
\end{equation}
in which the integration constant $C$ is determined by initial conditions. Although the form of Eq.~(\ref{eq:genform}) is more complex than Eq.~(\ref{ne_soln}), the conclusions are nevertheless the same: the reaction yield as a function of time is different from the uniform case (although, of course, the final yield once the reactants have been exhausted is always the same).
\section{Discussion}
We have shown here that for simple binary interactions leading to species abundance changes, gas kinetic theory  shows that the distribution function of the interacting species is altered, on short timescales, by their participation in the reactions. The result is that the mean energy associated with the reactants can change, influencing the progress of the reaction itself. For the idealised situation in which the species start off as Maxwellians, this feedback mechanism changes the temperature of the reactants (particularly the electrons) and can cause the reactant yield to alter significantly. If such calculations are carried out in the fluid limit using rate equations and rate constants, the latter can be shown to be subject to measurable changes even in the fluid limit.

In the zeroth-order fluid description, governed by rate equations, the species densities evolve without a self-consistent impact on the kinetic distribution of reactants (and therefore on the rate at which the reactions proceed). This article illustrates the risk in such an approach, but suggests a possible pragmatic way forward: rather than a full gas-kinetic simulation for at least the electrons, it may be possible to vary the rate coefficients in a way that reflects the underlying kinetic physics in the fluid limit.

In a more realistic situation, there will be multiple reactions underway simultaneously, each with a different set of cross-sections (in the kinetic limit) or rate constants (in the fluid limit) and each having a unique evolution under the changing distribution function.

Reaction-rate uncertainties due to non-equilibrium effects in plasmas have been postulated before 
\citep[for example][]{kruger} and demonstrated in experiments \cite{krugerexpt}, and there are numerical simulations of reaction rates via kinetic PIC codes for RF plasmas  \cite{tong} that show the limitations of depending on the chemical kinetic fluid approximations.

In practice, yield differences are apparent only for integration times of a few $\beta$; given that the typical electron number density in discharges is $10^{14-17}$m$^{-3}$, and that rate constants tend to have values in the range $10^{-15}-10^{-17}$m$^3$s$^{-1}$, then signficant drifts will occur over periods longer than $100\,$ms. This is too long for electronegative-type instabilities Eq.~\cite{corr03, lieb2, craigm}, but is less than the typical time for ozone destruction Eq.~\cite{chen}, for example. Hence ozone creation and destruction is a good test of the reasoning in this article, and there are several examples in the literature \cite{chen, soria, yanalla, mason} in which experiment and modelling are directly compared. The simulations are all based on the hydrodynamical rate equations, similar to Eq.~(\ref{reactionset}), though more extensive. In general, the qualitative agreement is very good, but the quantitative numerical predictions are not always reliable and accurate, with the calculated yields diverging from the experimental measurements significantly. Had the numerical simulations all been exact, then there would be no room for speculation that drift in the rate expressions may be relevant.

The simple fluid example given in this article is purely illustrative; in a real plasma there will be many such competing reactions. However, any temperature drift resulting from binary interactions will impact on species numbers across all such reactions. In the fluid approximation, the consequences for the rate equations are significant due to the rate of change of any one species being proportional to the product of the participating species number densities and the effective rate constant at that time. Of course, if there are no relaxation processes available to maintain the initial form of the distributions in all species, then even allowing for a time evolution of the rate constant will not be accurate enough to model reaction yields.

\acknowledgments  We are grateful to the following funding bodies for their financial support: Leverhulme Trust (LFAT fellowship); EPSRC for studentship funding (CSM); STFC for rolling grant support (HEP under ST/F002149/1). It is also a pleasure to acknowledge stimulating discussions with
Profs E W Laing and J C Brown.


%

\begin{description}
  \item[] 
  \item[ ]
  \item[ ]
\end{description}

\appendix
\section{Key integrals}
The following integrals are central to the results of this article, and are gathered together here for convenience.
\begin{equation}
\int_{-\infty}^{\infty}\exp(-ax^2)\,\mbox{d}x = \sqrt{\pi/a},\; a>0
\end{equation}
\begin{equation}
\int_0^{\infty}x\exp(-ax^2)\,\mbox{d}x = \case{1}{2}a^{-1},\; a>0
\end{equation}
\begin{eqnarray}
\int_{-\infty}^{\infty}|x-y|\exp(-ax^2)\,\mbox{d}x = ~~~~~~~~~~~~~~~~~~~\nonumber
\end{eqnarray}
\vspace{-0.5cm}
\begin{eqnarray}
~~~~~~\sqrt{\pi/a}|y|\,\mbox{erf}(\sqrt{a}|y|)+\exp(-ay^2)/a,\; a>0
\end{eqnarray}
\begin{equation}
\int_0^{\infty}x\exp(-ax^2)\,\mbox{erf}(bx)\,\mbox{d}x=\frac{b}{2a\sqrt{a+b^2}},\; a,b>0
\end{equation}

\pagestyle{empty}\end{document}